\documentclass[10pt,a4]{article}

\begin{document}

\begin{center}
{\large\bf
 Determination of constants of Standard Model and some

generalized models
\footnote{Talk at the International
Conference on Gravitation, Cosmology and Astrophysics

dedicated to 90th anniversary of K.P.Staniukovich, March 2-6, 2006, Moscow, Russia}
}

\medskip

{\bf V.V. Khruschov}\footnote{e-mail: khru@imp.kiae.ru}
\end{center}
\medskip
Russian Research Centre "Kurchatov Institute", Kurchatov Sq.,
123182 Moscow\\
and\\
Centre for Gravitation and Fundamental Metrology, VNIIMS, Moscow

\smallskip

\begin{abstract}
Methods of determination of constants of the Standard Model
are considered. The constants values obtained now are presented and
experiments for improving some  values are pointed out. A few
possible generalized models are considered together with their
groups of gauge and kinematical symmetries.
\end{abstract}

Determination of precise values of the Standard Model (SM) constants
is the present-day issue of the modern physics. SM is the quantum
gauge theory of strong, electromagnetic and weak interactions and
include in the quantum electrodynamics (QED) \cite{berest}, the
quantum chromodynamics (QCD) \cite{indura} and the theory of weak
interactions (WIT) \cite{li}. In the SM framework WIT and QED are
united in the single gauge theory of electroweak interactions (EWIT)
-- the Glashow-Weinberg-Salam theory \cite{wein}. Values of large
number of physical processes characteristics have been evaluated in
SM, many of them have been confirmed by experimental data.
Notwithstanding significant achievements of SM most of researchers
considered and consider at present the SM as the transition theory
along the path to a creation of a more perfect and universal theory.
An argument in favor of this opinion is a large number of  SM
constants, which are the external parameters of the theory, and
there are no explanations as the existing values as some connections
between them. The possibility of time variations of SM constants is
also very intriguing issue and there are a number of estimations of
such variations, for  instance, in the framework of  grand
unification theories (GUT) \cite{uzan,mel,ivas,kon_mel}. Moreover
recently the experimental data appear, which confirm the necessity
for a SM extension. The discoveries of neutrino oscillations and
nonzero neutrino masses, as well as  new physical substances - the
dark matter and the dark energy - are facts of this kind
\cite{bine}. It is necessary to point out that in the SM framework
several problems remain open, such as the spontaneous breaking of
chiral symmetry, the confinement of quarks and gluons, the
experimental verification of existence of Higgs particles
\cite{altar}. Let us consider the set of SM constants, some methods
of  determinations and  improving their values. A few examples  
of generalized models  are cited, these models can be used 
for a SM extension, in search of new physics and a construction 
of a unified theory.

It is known,  that at present the determination of the constants of 
electromagnetic, strong and weak interactions are carried out 
using theoretical evaluations of quantum effects in the SM framework,
accumulation of experimental data with modern measurement equipments, 
processing and comparisons of theoretical and experimental results obtained
\cite{mohr,eidel}. The main part of the data concerning constants of 
electromagnetic, strong and weak interactions was found with the help of 
powerful accelerations and detectors of elementary particles. 
For carrying out SM model calculations one needs to determine
19 parameters or SM model constants. These parameters are as follows:
the fine structure constant $\alpha$, the Fermi constant $G_F$, 
the $Z$ bozon mass $M_Z$, the Higgs bozon mass $M_H$, the mass values of 
three leptons and six quarks, four parameters of quarks mixing, the value of 
the QCD CP violation phase and the strong interaction coupling $\alpha_s(M_Z)$
at the $Z$ bozon mass. It should be noted that the QCD CPV phase value is very
small ($< 10^{-9}$ \cite{eidel}), so as usual it is suppose that its value
is equal to zero. The Higgs bozon mass is unknown parameter, because of
at present the Higgs bozon is undiscovered. However the  Higgs bozon 
search is proposed to carry out at the new accelerators such as the LHC
at CERN.

For matching results of quantum and classical measurements the Thirring's
theorem is important, according to which   quantum and classical formulae
for a foton scattering on a electron must coinside each other \cite{thirr}.
The value of the fine structure constant $\alpha$ is determined with the best 
accuracy from measurements of the anomalous magnetic moments of $e^{\pm}$,
its value at low energies is equal to $\alpha = 1/137.03599911(46)$.
When the energy is increased the $\alpha$ value is also increased and
at the $Z$ bozon mass its value  is equal to $\alpha(M_Z) = 1/127.918(18)$
\cite{eidel}. The Fermi constant $G_F$ is determined using the  lifetime
of the $\mu$ meson $G_F = 1.16637(1)\times10^{-5}$, the mass value 
of $Z$ bozon is determined using the resonance line shape, obtained 
at LEP 1 accelerator $ M_Z = 91.1876(21) Ē'$ \cite{eidel}.
The lepton masses are the following: $m_e= 0.51099892(4)$
MeV, $m_{\mu}= 105.658369(9)$ MeV, $m_{\tau}=1776.99(29)$ MeV 
\cite{mohr,eidel}. It should be noted, that the mass values of
electron and muon are measured with best accuracy in the atomic mass units 
$u$ (for instance, $m_e= 5.4857990945(24)\times10^{-4} u $ 
has been obtained through the comparison of cyclotron frequencies of electons
and a single carbon ion in a Penning trap \cite{farnham}). This fact counts
in favour of the way for replacing the kilogram prototype with a new standard,
which will be based on two fundamental physical constants, namely, the 
Avogadro constant and the atomic mass unit (see, e.g. \cite{mills,kon}).

Lepton masses, in principle, can be determined in free states of leptons
and they have classical limits in distinction to quark masses, which have
no any measurement procedure of such type. Quark masses cannot be normalized
to  "physical mass values" at some scale. They cannot be determined 
unambiguously because of their values depend on a renormalization 
scheme  and a renormalization point. The renormalization scale dependent
$m(\mu)$ values must satisfy the following equation \cite{indura,eidel}:

\begin{equation}
\mu^2 dm(\mu)/d\mu^2 = -\gamma(\alpha_s(\mu))m(\mu),                   
\end{equation}
where $\gamma$ is an anomalous dimension, which is known to fourth order of
the QCD perturbation theory, with
\begin{equation}
\gamma(\alpha_s(\mu)) = \sum_1^{\infty}\gamma_r(\alpha_s/4\pi)^r,
\quad \gamma_1=4, \quad\gamma_2 = 202/3 - 20n_f/9, \quad ...,         
\end{equation}

\noindent $n_f$  is a number of active quarks in a process under consideration,
with masses met the condition $m < \mu$. 
So, quarks become lighter when  a magnitude of an momentum transferred increase.
The quark mass values in $\overline{MS}$ scheme at $\approx 2 GeV$ are as 
follows: $ m_u =
1.5\div4 MeV$, $m_d = 4\div8 MeV$, $m_s = 80\div130 MeV$, 
$m_c=1.15\div 1.35 GeV$, $m_b = 4.1\div4.4 GeV$ \cite{eidel}.

A dependence of the QCD constant $\alpha_s=g_s^2/4\pi$, where $g_s$
is the coupling constant between quarks and gluons, can be evaluated in the 
QCD perturbation theory framework and subsequently can be measured 
experimentally in processes, when magnitudes of  momenta transferred are 
larger than the QCD constant $\Lambda_{QCD}$, which depends on the 
renormalization point and the number of active quarks in the process  
considered ($\Lambda_{QCD}$  is of order of several hundreds MeV). 
$\alpha_s$ values are determined  with data processing from deep-inelastic 
lepton-hadron scattering, high energy hadron-hadron scattering, 
electron-positron scattering, heavy particles decays, $\mu$ meson decays 
\cite{eidel}. The main contribution in the total uncertainty
for $\alpha_s$ values is due to systematic uncertainties, which are, as a rule, 
theoretical ones. The renormalization scale dependence of 
the effective QCD coupling $\alpha_s$ is governed by the equation:
\begin{equation}
\mu\partial\alpha_s(\mu)/\partial\mu = 2\beta(\alpha_s) = -\beta_0/2\pi\cdot\alpha_s^2 
-\beta_1/4\pi^2\cdot\alpha_s^3 -\beta_2/64\pi^3\cdot\alpha_s^4 - \cdots,
\end{equation}

\noindent where $\beta_0 = 11 - 2n_f/3$, $\beta_1 = 51-19n_f/3$,
$\beta_3 = 2857-5033n_f/9 + 325n_f^2/27$, $m_{n_f} <\mu$.
In the one-loop approximation the dependence of $\alpha_s$ 
from the magnitude  of  momentum transferred has the form:
\begin{equation}
\alpha_s(Q^2)
=\alpha _s(\mu^2)/(1+\alpha_s (\mu^2)/4\pi\beta_0
ln(Q^2/\mu^2)),
\end{equation}

\noindent with in  the QED $\beta_0 = -4/3$, in the EWIT $\beta_0 = 10/3$, at
$ n_f =6$, in the QCD $\beta_0 = 7$, at $ n_f =6$. Thus, 
the  QED coupling $\alpha$ is growing when $Q^2$ is increased, while
the WIT coupling $\alpha_w$ and
the  QCD coupling $\alpha_s$ are vanishing with $Q^2$ in the asymptotic
freedom region. The commonly used scale for a tabulation of SM constants
is the Z bozon mass. To cite the $\alpha_s$ value at this scale \cite{eidel}:
\begin{equation}
\alpha_s(M_Z) = 0.120\pm 0.002(exp)\pm 0.004(thr)        
\end{equation}

As is evident from the foregoing there is the well developed procedure
for determination the strong coupling constant $\alpha_s$ at large magnitudes
of $Q^2$, but when a momentum magnitude is in the neighbourhood of a quark
mass it is necessary to take into account threshold corrections.
For  determination of $\alpha_s(Q^2)$ below and 
above the threshold the  following condition is used \cite{eidel}:
\begin{equation}
\alpha_s(n_f , M_{n_f}) =\alpha_s(n_f -1, M_{n_f}) - 
11\alpha_s^3(n_f -1, M_{n_f})/72\pi^2 + O(\alpha_s^4(n_f -1, M_{n_f}))       
\end{equation}

It should be noted, that the $\Lambda_{QCD}$ value in its turn is dependent on
a number of active quarks $n_f$, that is, $\Lambda_{QCD}$ $\to$ 
$\Lambda_{QCD}^{n_f}$, with  a $\Lambda_{QCD}^{n_f}$ value become less
when a number of active quarks increases, for instance, $\Lambda_{QCD}^{5} =$
$217\pm25 MeV$ at $n_f=5$ \cite{eidel}. When momentum magnitudes are comparable
with $\Lambda_{QCD}$, it is essential to take into account nonperturbative 
contributions, which at $Q^2 \ll \Lambda_{QCD}$ go dominant. This momentum 
domain is reffered to as the QCD nonperturbative domain or the domain of 
confinement of quarks and gluons. There is a widespread opinion, that the 
confinement of quarks and gluons can be established rigorously in the QCD 
framework. Is this is the case, a exact dependence should take place, which 
relates $\Lambda_{QCD}$ to the characteristic parameter of strong interactions
 in the  nonperturbative domain -- the  string tension $\sigma$. However, 
in the case if the confinement cannot be demostrated in the QCD framework, 
there has to add the number of strong interaction parameters and to introduce,
 at the least, one more parameter -- the string tension. From this standpoint 
of some interest is investigations concerning with the generalization of the 
space-time Poincare symmetry of the QCD in the confinement domain to the 
inhomogeneous pseudounitary symmetry  IU(3,1) \cite{kh}, along with the 
universality of the confinement potential verified in the framework of the 
relativistic model of quasi independent quarks \cite{kh_s_s}. In the framework
of this model the value of the string tension was found to be 
$\sigma=0.20\pm0.02$ $GeV^2$ for light and heavy quarks, furthermore
new fundamental constants for the confinement domain with dimensionalities
of  mass and length can be associated with the universal  string tension:
$m_c = 0.45\pm0.02$ $GeV$, $l_c = 0.44\pm0.02$ $Fm$. The number of SM parameters
 also increases, when one go from the quantum field theory  in the Minkowski
space-time to a quantum field theory  in a curved space-time or
a quantum phase space. We shall exemplify below the commutation relations
between the observables of the quantum field theory  in the
 quantum phase space, which have been obtained subject to the Lorentz 
invariance and the principle of correspondence with the conventional
commutation relations after going to the limit: $H\to\infty$, $L\to\infty$,
$M\to\infty$ \cite{kh_lez}.
$$
\hspace{1.2cm}[F_{ij},F_{kl}] = if(g_{jk}F_{il} - g_{ik}F_{jl} + 
g_{il}F_{jk} - g_{jl}F-{ik} ),
$$
$$
\hspace{-2.5cm}[p_i,x_j] = if(g_{ij} I + F_{ij}/H),
$$
$$
\hspace{-3.7cm}[p_i,p_j] = ifF_{ij}/L^2,
$$
$$
\hspace{-3.4cm}[x_i,x_j] = ifF_{ij}/M^2,
$$
\begin{equation}
\hspace{-2.4cm}[p_i,I] = if(x_i/L^2  - p_i/H),
\end{equation}
$$
\hspace{-2.2cm}[x_i,I] = if(x_i/H  - p_i/M^2 ),
$$
$$
\hspace{-2.2cm}[F_{ij},p_k] = if(g_{jk}p_i - g_{ik}p_j ),
$$
$$
\hspace{-2.1cm}[F_{ij},x_k] = if(g_{jk}x_i - g_{ik}x_j ),
$$
$$
\hspace{-4.8cm}[F_{ij},I] = 0
$$

In particular limiting cases the algebra of observables given above transforms
to algebras obtained and considered  by a number of authors \cite{snyder,yang,
gol_,kady,toller}. For example, at $H\to\infty$, $L\to\infty$
one can derive the Snyder theory with the fundamental length 
$a=f/M$ \cite{snyder}. At $H\to\infty$ there is the Yang theory
with the small fundamental length $a$ and the large fundamental length $R$
\cite{yang}. In our designations $R=L$ and one can say about the 
 fundamental mass $m=f/L$. In general case the algebra presented
contains the constant $L$ with dimensions of length, the constant $M$
with dimensions of mass and two constants $f$ and $H$ with dimensions of 
action. The constant $f$  must convert to the Plank constant $\hbar$ 
subject to the fulfilment of the principle of correspondence, whereas
the constant $H$ conceptually can be  different from  $\hbar$ and in this
case its presence causes the theory to violate the $CP$ invariance.
Some consequences of the theory, which incorporates the new constant $H$,
have been deduced in Ref. \cite{leznov}. General mathematical properties
of algebras  cited above and their conceivable physical interpretations
are considered  in the work \cite{chryss}.

In addition to the confinement problem in the SM framework
there are problems concerning  spontaneous breaking of chiral
invariance,  existing of the standard Higgs bozon, as well as
a verification of a time dependence of SM constants and
decreasing a great deal of them. Certain of the problems
should receive their solutions in the framework of a more general
theory than the SM \cite{eidel,sandvik}. In doing so the 
$SU(3)_c \times SU(2)_L \times U(1)_Y$ gauge symmetry
group of SM go over into  a more extended gauge group of GUT, possessed,
as a rule, a more simple structure. Pioneering GUT theories
were based  on gauge groups like as  $SU(4)_{ec}\times SU(2)_L\times
SU(2)_R$ \cite{pati}, $SU(5)$ \cite{geor_gla}, and $SO(10)$
\cite{fritzsch,georgi}. At present of main interest is the $SO(10)$
model with the following pattern of spontaneous breaking of 
the gauge symmetry: $SO(10) \to$ $SU(4)_{ec}\times SU(2)_L\times
SU(2)_R \to$ $SU(3)_c \times SU(2)_L \times U(1)_Y$.
This pattern does not contradict the existing restrictions associated
with the value of the Weinberg angle, the lifetime of the proton
and  neutrino masses \cite{eidel}. Of some interest are  also
supersymmetric GUT within the context of which the experimental restrictions
can be satisfied, however the supersymmetric GUT 
have diversified assortments of new particles as compared with 
nonsupersymmetric GUT.  In order to one has no contratictions with
experimental data, these new particles must acquire great mass values.
 
Nowadays a possibility of time and space variations of fundamental physical 
constants are being studied intensively as in the GUT context, as on the
phenomenological level. Results of theoretical and experimental investigations 
alongthese lines are represented in Refs. \cite{uzan,mel,ivas,kon_mel,
sandvik,peres}. Limitations for time variations of constant of
fundamental interactions can be obtained using  data concerning
the nucleosynthesis of elements during the Big Bang, electromagnetic 
spectra of quasars and the analysis of elements from the native 
nuclear reactor in Oklo. As the most probable experimentally variations
are for the fine structure constant $\alpha$, we know that in the GUT
framework time variations of $\alpha$ are related with time variations
of other constants including mass values of fundamental particles
(see, e.g. \cite{fr}). In Refs. \cite{lang,flambaum} it was observed,
that $\delta\Lambda_{QCD}/\Lambda_{QCD}$$\approx 34\delta\alpha/\alpha$, 
$\delta m_q/ m_q $$\approx 70\delta \alpha/\alpha$, 
$\delta(m_q /\Lambda_{QCD})/
(m_q /\Lambda_{QCD})$$\approx 35 \delta\alpha/\alpha$. These estimations
indicate that the time variations are more evident through strong
interaction effects, what should be taken into account in preparations
for future experiments. Moreover, contributions of strong interaction effects
are very important for explanations of experimental data related to the 
anomalous magnetic moment of the muon \cite{kataev} and the hyperfine splitting
of energy levels of the positronium \cite{penin}.

It should be particularly emphasized that at the present time
the experimental facts concerning the new phenomena beyond the SM 
have been established. In the Universe the dark matter and the dark energy 
are discovered and perhaps new particles and interactions demand for
explanations of their existence. Owing to the high precision and
long-run experiments it has been found, that neutrino have nonzero masses
and neutrino with different lepton numbers mix together what leads to
neutrino oscillations. These facts are of primary interest for a description of
neutrino properties, as well as for the development of a theory, which
generalize the SM and make it possible to evaluate some SM constants.

\end{document}